\documentclass[5p]{elsarticle}
\usepackage{wrapfig}
\usepackage{textcomp}
\usepackage{hyperref}
\usepackage{lineno}\modulolinenumbers[5]

\journal{Fusion Engineering and Design}

\begin{document}

\begin{frontmatter}

\title{SMITER: A field-line tracing environment for ITER}

\author[lecad]{L. Kos}
\ead{leon.kos@lecad.fs.uni-lj.si}
\author[iter]{R. A. Pitts}
\author[lecad]{G. Simi\v{c}}
\author[lecad]{M. Brank}
\author[iter]{H. Anand}
\author[culham]{W. Arter}
\address[lecad]{LECAD laboratory, Mech.Eng., University of Ljubljana, Aškerčeva 6, 1000 Ljubljana, Slovenia}
\address[iter]{ITER Organization, Route de Vinon-sur-Verdon, CS 90 046, 13067 St. Paul Lez Durance Cedex, France}
\address[culham]{Culham Centre for Fusion Energy, Culham Science Centre, Abingdon, OX14 3DB, Oxfordshire, U.K.}

\begin{abstract}
Built around the SMARDDA modules for magnetic field-line tracing [IEEE Tr.~Plasma~Sc. 42 (2014) 1932], the SMITER code package (SMARDDA for ITER) is a new graphical user interface (GUI) framework for power deposition mapping on tokamak plasma-facing components (PFC) in the full 3-D CAD geometry of the machine, taking as input a user-defined specification for parallel heat flux in the scrape-off layer (SOL) and a description of the equilibrium magnetic flux. The software package provides CAD model import and integration with the ITER Integrated Modelling and Analysis Suite (IMAS), parametric CAD components catalogue and modelling, CAD de-featuring for PFC surface extraction, meshing, visualization (using an integrated ParaView module), Python scripting and batch processing, storage in hierarchical data files, with several simulation cases in one study running in parallel and using message passing interface (MPI) for code speed-up. An integrated ParaView module can combine CAD geometry, magnetic field equilibrium, meshes and results for detailed setup analysis and a module is under development for full finite element computation of surface temperatures resulting from the power deposition patterns on 3-D PFCs. The code package has been developed for ITER, but can be deployed for similar modelling of any tokamak. This paper presents and discusses key features of this field-line tracing environment, demonstrates benchmarking against existing field-line tracing code and provides specific examples of power deposition mapping in ITER for different plasma configurations.
\end{abstract}


\end{frontmatter}


\section{Introduction}

The ITER plasma-facing components (PFC) are now fully designed and procurement is underway. A key utility in such design is field-line tracing for different magnetic field equilibria which allows the definition of component front surface shaping. On ITER, this design phase has deployed both analytic theory~\cite{Stangeby} and the field-line tracing code PFCFLUX~\cite{Firdaouss}. Attention is now turning towards the critical issue of management and control of PFC heat fluxes in an actively cooled device such as ITER. To facilitate the development of control algorithms, particularly for protection of the beryllium main chamber first wall panels ~\cite{Anand}, a new field-line tracing environment, SMITER, has been developed, featuring a sophisticated graphical user interface (GUI) that uses the SMARDDA ~\cite{Arter} kernel and has been thoroughly benchmarked against PFCFLUX for specific cases of first wall panel and divertor target loading~\cite{Anand}. 
 
The main difficulty in field-line simulations is run-case setup which usually involves different commercial packages starting with CAD models mostly targeted at representation and manufacture. Simulations require defeatured models resulting in higher quality meshing and consequently fewer numerical problems in simulations. 

Field-line traces require only an exact PFC surface representation and a magnetic equilibrium as input. However, ``exact'' representation of the PFC surface using triangular grids requires a large number of triangles which are limited by memory and processor capabilities. Moreover, the magnetic equilibrium input is often prescribed on a low resolution grid which requires ``smooth'' interpolation of derivatives for every point of the field-line trace on a given magnetic field line. Pseudo-time integration of each field-line needs to be quite precise since the field-line angles of incidence on PFCs are generally low in tokamaks owing to the strong toroidal field in comparison to the poloidal field produced by the plasma current. In turn, these low incidence angles mean that PFC surfaces are typically shaped to ensure that leading edges arising as a consequence of assembly and manufacturing tolerance cannot be accessed by field lines on which high plasma power flux densities can flow. Such leading edge protection is even more important on an actively cooled device like ITER.

\subsection{Field-line tracing}

The approach is to use the standard field-line equation
\begin{equation}
\label{eq:fieldline}
\dot \mathbf{x} = d\mathbf{x}/dt=\mathbf{B}(\mathbf{x}),
\end{equation}
where the dot denotes differentiation with respect to pseudo-time $t$ measured along the fieldline~\cite{Arter}, assuming the guiding centre approximation (neglecting the ion Larmor radius). The term "pseudo-time" refers to solving the steady-state solution of differential equation~(\ref{eq:fieldline}) by prescribing small enough particle time step to maintain fieldline accuracy. 

The magnetic field $\mathbf{B}$ in SMITER can be fully 3-D or can be assumed
axisymmetric, independent of toroidal angle $\phi$, and thus
described by a (poloidal) flux function $\psi$, together with a toroidal
component specified by the flux function $I(\psi)$. Pseudo-time $t$ can be
considered as field-line length.
In cylindrical coordinates the magnetic field components can be written as
\begin{equation}
  \label{eq:cylindrical}
  B_R=-\frac{1}{R}\frac{\partial \psi}{\partial Z},~~
  B_T=\frac{I(\psi)}{R},~~
  B_Z=\frac{1}{R}\frac{\partial\psi}{\partial R}~,
\end{equation}
where $B_T$ is the toroidal component of the field, directed along the $\phi$ coordinate. In flux coordinates ($\psi$, $\theta$, $\phi$) the field-line equation simplifies to ${d\theta}/{d\phi}=(1/I)R/J(\psi, \theta)$, where J($\psi$, $\theta$) is the Jacobian of the mapping transformation. However, this simplified equation in flux coordinates can only be used where simple 2D mapping from/to global space exists thus field lines must not intercept X-points. Generally, this means that for diverted configurations, the standard field-line equation in a ``global'' calculation must intercept one of a set of surfaces specified by the ``termplane'' data structure to count as illuminated, whereas for limiter cases a ``local'' escape calculation implies that the triangle is illuminated if the field-line starting from it leaves the computational domain without intersecting another triangle in the shadowing set.

\subsection{Power deposition model}
\begin{wrapfigure}[23]{r}{0.38\columnwidth}
  \begin{center}
  \vspace{-25pt}
    \includegraphics[width=\linewidth]
    {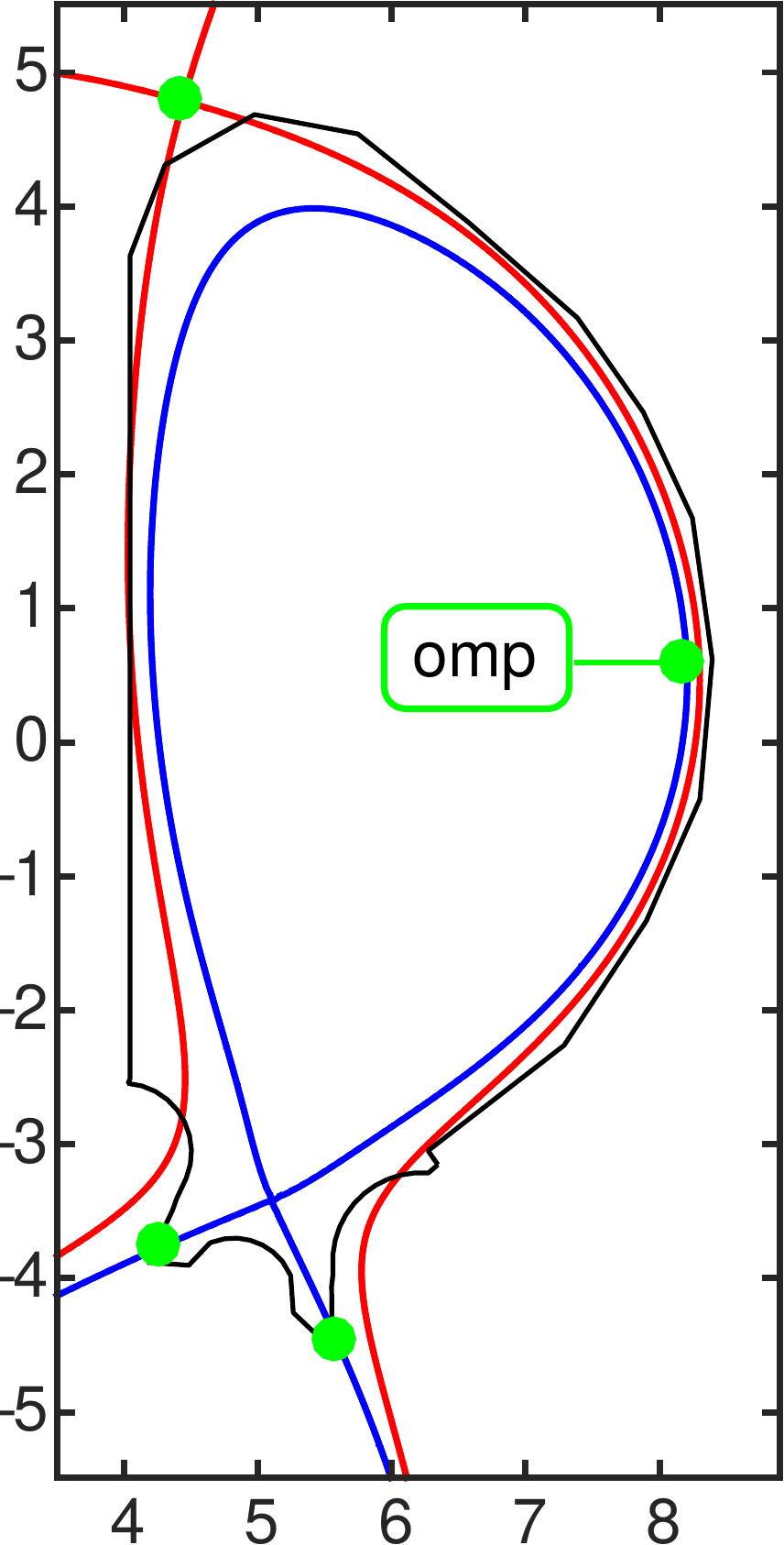}
  \end{center}
  \caption{Illustrating the definition of primary (blue) and secondary (red) separatrices for  the ITER baseline burning plasma magnetic equilibrium.}
  \label{fig:power_dep_model}
\end{wrapfigure}
SMITER maps profiles of scrape-off layer (SOL) heat flux density flowing parallel to the magnetic field lines $q_{||}$ onto PFC surfaces.
 
Generally, the heat flux parallel to the magnetic field lines is assumed to fall off exponentially radially into the SOL from the primary separatrix (diverted configurations) or from the last closed flux surface (LCFS) in limiter configurations: 
\begin{equation}
  \label{eq:q_par}
  \!\!\!\!\!\!\!\!\!\!\!%
  q_{||}(R)=q_{||omp}e^{-(R-R_m)/\lambda_q}~,
\end{equation}
where $q_{||omp}$ is the parallel heat flow at the outer midplane (omp) separatrix (m) or LCFS (blue line in Fig.~\ref{fig:power_dep_model}), with radius $R_m$ and $\lambda_q$ is the characteristic width of this exponential decrease (the distance over which $q_{||omp}$ falls to $1/e$ of its value).

The Z-position of the omp is defined by the centre of the magnetic equilibrium flux function $\psi$. The second separatrix, applicable to diverted configurations (red line in Fig.~\ref{fig:power_dep_model}), is defined as the first flux surface to contact any part of the wall contour (black lines in Fig.~\ref{fig:power_dep_model}) before reaching the divertor targets. In the case of the equilibrium in Fig.~\ref{fig:power_dep_model}, the second separatrix contacts the first wall at the top of the chamber and the upper part of the outer divertor baffle. 

It is convenient to use flux coordinates $\psi$ in Eq.~(\ref{eq:q_par}) by introducing approximation for
\begin{equation}
   \!\!\!\!\!\!\!\!\!\!\!%
  \Delta R = R-R_m \approx \frac{\partial R}{\partial\psi}\Delta\psi 
  = \frac{\partial R}{\partial\psi} (\psi - \psi_m)  = \frac{\psi-\psi_m}{R_m B_{pm}}~,
\end{equation}
where the poloidal component of $\mathbf{B}$ at the midplane is given by $B_{pm} = (1/R_m)(\partial\psi/\partial R)$ [cf. Eqs.~(\ref{eq:cylindrical})] assuming radial component of the field is small.

Power deposition $q_{PFC} \propto \mathbf{B}\cdot\mathbf{n}$, with surface normal $\mathbf{n}$, assumes flux conservation $\nabla\cdot\mathbf B = 0$ that results (see Ref.~\cite{Arter}) in 
\begin{equation}
\!\!\!\!\!\!\!\!\!\!\!%
  q_{PFC}(\psi) = \frac{P_{SOL}}{4\pi R_m\lambda_q B_{pm}}\mathbf{B}\cdot\mathbf{n}
  \exp\left(-\frac{\psi-\psi_m}{R_m B_{pm}\lambda_q}\right)~,
\end{equation}
where $P_{SOL}$ is the power flowing into the SOL across the separatrix/LCFS (assumed to enter the SOL at the omp). SMITER can also include alternative models for the radial profile of parallel heat flow, such as double exponentials~\cite{Kocan}, the divertor spreading formula defined in Ref.~\cite{Arter} or more complex expressions which properly take into account the variation of field line connection length in the SOL (see for example the application of the ELM parallel loss model~\cite{Kocan,Firdaouss}).

\section{SMITER simulation principles}

\begin{figure*}[t]
  \begin{center}
    \includegraphics[width=1.0\linewidth]{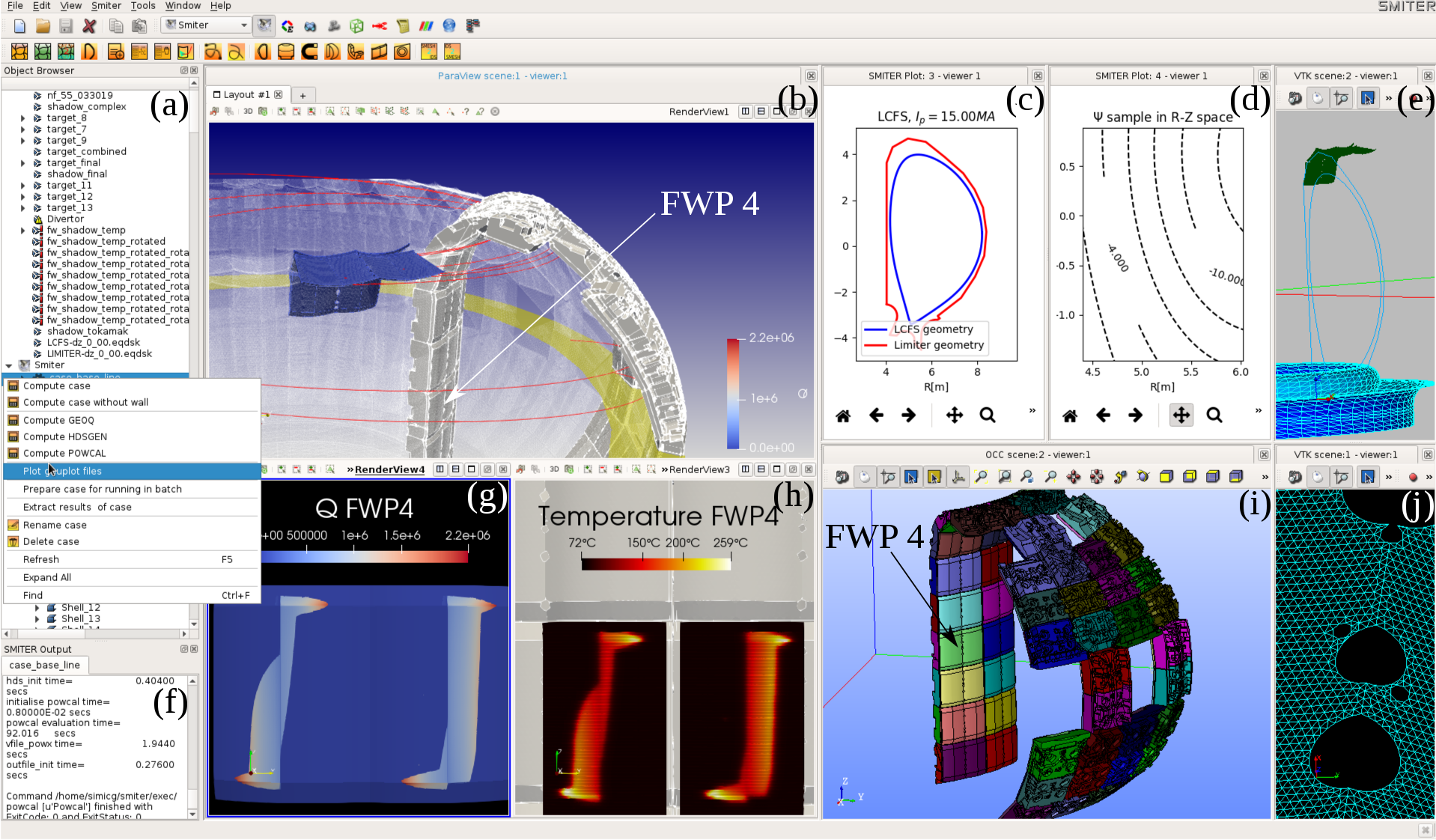}
  \end{center}
  \caption{Illustration of the SMITER graphical user interface for ITER. The code allows several
    run-cases in one study (a) to be run (f) in parallel on a compute cluster. ParaView window (b) shows resulting target top panels (blue), selected characteristic field lines (red), omp disk (yellow), a complete blanket sector CAD model from the Geometry module and a shadow mesh from the Meshing module (grey) augmented for an overall evaluation of run-case setup. The magnetic equilibrium with LCFS and Limiter/Wall geometry (c) and other details can be further analyzed with built-in 2D and 3D plots such as flux function detail (d). Triangular meshes (j) are the main run-case geometry setup (e) that are directly imported or meshed from CAD models (i) defeatured to retain only the required PFC surfaces for meshing with different algorithms and hypotheses. Resulting heat fluxes (g) on the ITER first wall panel number 4 (FWP4) can be further processed to get temperatures (h) using FEM thermal models of normal heat flux (NHF) or enhanced heat flux (EHF) cooling sub-structures.
    }
  \label{fig:SMITER_GUI}
\end{figure*}

To map a specified omp power deposition radial profile on PFCs, magnetic field lines on flux surfaces within the magnetic equilibrium need to be followed in 3D space until they intersect a solid surface. The PFC geometry is obtained from CAD models of the structure converted to a high precision triangular mesh. In practice, field lines are followed backwards from the surface in question, with proper mapping of the heat flux profile specified in the free SOL.

The field-line tracing must take into the account the neighbouring structures around the object of interest in order to ensure that the field lines are not intersected by other solid surfaces. If this does occur, zero heat flux density is associated with the point on the surface at which the field line originated. 

SMITER algorithms are programmed for limiter and divertor cases.  The geometry for both of them is separated into two types, taking into account of the wetted and shadowed
regions.

The SMITER framework is composed of several modules for pre- and post- processing. Pre-processing provides transformation of input CAD surfaces into meshes required by the main SMITER module which performs the field line trace. Post-processing allows analysis of the results with a ParaView visualization module and exporting data to other formats.

Data workflow between modules is based on Common Object Request Broker Architecture (CORBA) inter-process communication which provides methods and objects for scripting in Python programming language. The SMITER module is a main CORBA component that runs several external FORTRAN codes. Meshing of the CAD model built from curves and surfaces which are precisely defined by the bicubic splines can be performed within SMITER meshing module. Alternatively, externally generated triangular meshes can be imported/exported in Nastran/Patran format or from/to IMAS Wall Interface Data Structure (IDS). The magnetic equilibrium required by SMITER for the field line trace can be read as a standard equilibrium (EQDSK) file format (the output of the EFIT code widely used in the tokamak community) or imported from an IMAS equilibrium IDS. SMITER module runs external FORTRAN codes in the following order: 
\begin{enumerate}
\item The geometry and equilibrium processing code (GEOQ) is used to
analyse the flux function usually defined in the EQDSK file and
prepares the bicubic spline-interpolated flux function. 
\item The hybrid data structure generator code (HDSGEN) computes the multi-octree
hierarchical data structure (HDS)~\cite{Arter}, which is designed to accelerate the
computation of field line-triangle intersection.
\item The power calculator code (POWCAL) performs the power surface deposition calculation, following the field lines using the transformed geometry from GEOQ and the HDS from HDSGEN. POWCAL uses MPI parallelism to speed up field line tracing.
\end{enumerate}

\section{Graphical User Interface}

The SMITER graphical user interface (GUI) shown in Fig.~\ref{fig:SMITER_GUI} uses several modules (or components) which may be used independently and interoperably in the workflow based on the SALOME pre-post processing framework~\cite{salome}. Modules create data which is usually stored into a study that is saved into hierarchical data format (HDF) file for reuse. The geometry module (GEOM) provides basic CAD operations for creation, extraction, healing, and modification of CAD models. It can import CAD files in many different formats, including standard CAD formats for data exchange -- STEP and IGES and permitting the creation of geometrical and topological objects with different modelling operations.

A key element for mesh preparation is the generation of PFC surface models
from imported CAD models. GEOM provides explosion of the assembly into faces and grouping back into compounds for the building of a complete PFC model. The meshing module uses a set of meshing algorithms and their corresponding conditions (hypotheses) to generate  meshes from geometrical models created in or imported into GEOM. The main functionalities are creation of meshes, grouping, transformations, and mesh modifications. Meshing of CAD models can be modified by a density parameter change upon user request allowing gradual increase of precision during the preparation of cases.  ParaViS is a data analysis post-processing module which embeds ParaView visualization tool inside the SMITER GUI. Through the CORBA interface CAD models or meshes can be directly imported for compound representation as shown in Fig.~\ref{fig:SMITER_GUI}(b). For such visualization purposes the SMITER module includes parametric CAD component generation scripts with dialogs which build assemblies of cryostat, magnets, vacuum vessel, divertor, monoblocks, blanket, panels and other CAD modeling examples at different levels of detail for future expansion to other codes beyond field-line tracing. Additionally, SMITER includes ELMER finite elements modelling (FEM) module that enables calculation of stationary PFC surface temperature distributions. This module includes routines for thermal calculation including the full sub-structure of the temperature field resulting from a given heat flux density distribution on the top surface of the PFC.

\section{Benchmarks}
\begin{figure}[tb]
  \centering
  \includegraphics[width=1\columnwidth]{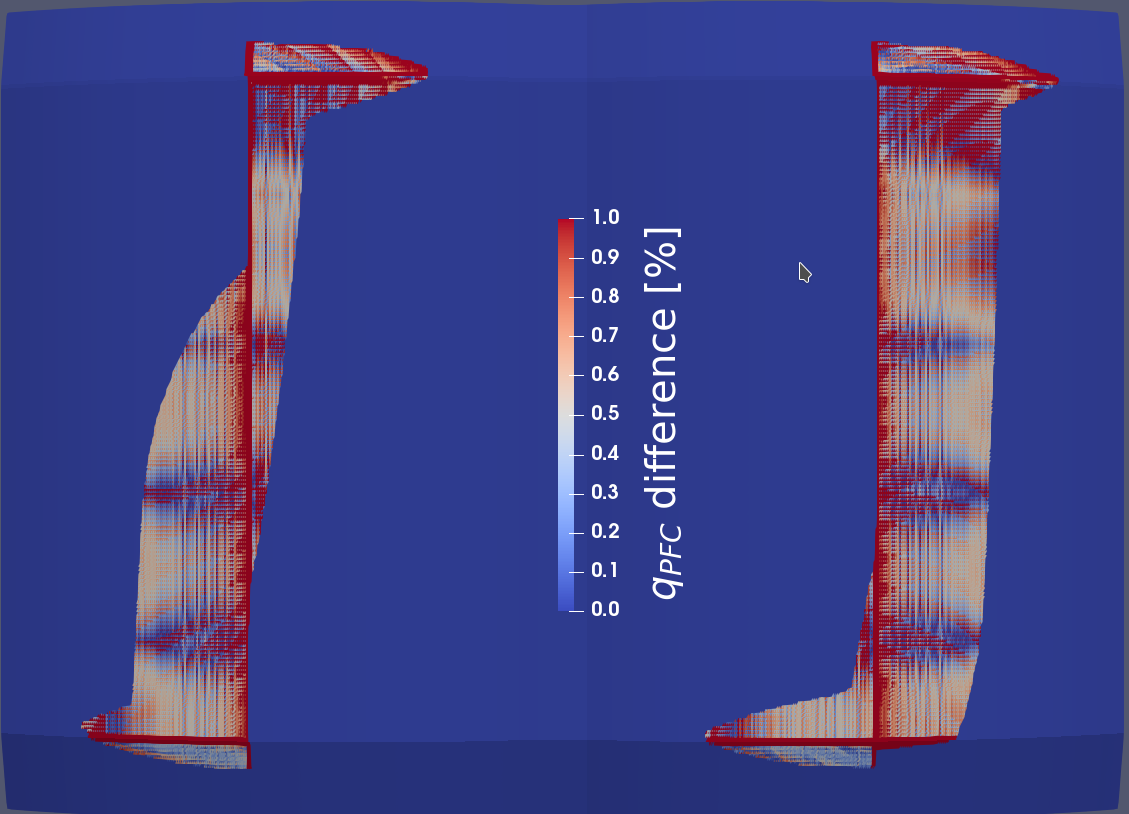}
  \caption{Heat flux relative difference between SMITER and PFCFLUX on FWP 4. Note that there are two wetted zones corresponding to the two apexes of the double winged panel design~\cite{Kocan}.}
  \label{fig:nf55-Q-difference}
\end{figure}

SMITER has been extensively verified and validated for both divertor and limiter configurations. The ensemble of these ready-to-run HDF studies are part of the ITER software distribution. Two particular examples are highlighted here for the purposes of demonstration using previous studies with the PFCFLUX~\cite{Firdaouss} field-line tracing code. The first is the case of an ITER inner wall limiter start-up equilibrium on FWP 4 examined with PFCFLUX in Ref.~\cite{Kocan}. Adopting the same case setup parameters (a single exponential omp $q_{||}$ profile with $\lambda_q$ = 50\,mm and the same $P_{SOL}$ - see Fig.~11(b) in Ref.~\cite{Kocan}) and identical meshes, the maximum heat load obtained with PFCFLUX is $q_{PFC} \sim 2.26\,MW/m^2$, while SMITER computes the maximum heat load of $q_{PFC} \sim 2.27\,MW/m^2$. The surface heat loads (Fig.~\ref{fig:nf55-Q-difference}) are within 1\,\% between the two codes. 
Subtle differences between the codes observed as red peaks correspond to sharp surface variations. Closer examination reveals red bands of a few triangle in width which are caused by PFCFLUX averaging surface normals from nodes while SMITER computes normals from a single triangle. Therefore, SMITER is more accurate on edges and has some oscillation of normals on smooth surfaces. More details can be found in the discussion of surface accuracy in Section II-C3 and Support Section II of Ref.~\cite{Arter}. Magnetic field interpolation differences are in the range of 0.006\,\% following the slow vertical variations observed in Fig.~\ref{fig:nf55-Q-difference}.

The second benchmark example is more complex and concerns the heat loads during the current quench (CQ) phase of an upward going vertical displacement event on ITER at baseline (15\,MA) performance. Here, the loads are mainly concentrated on FWP 8-10 and were first estimated using PFCFLUX to provide the estimates of the likely extent of CQ driven beryllium wall evaporation and melting first shown in Ref.~\cite{Lehnen}.   
\begin{figure*}[t]
  \centering
  \includegraphics[width=0.8\textwidth]{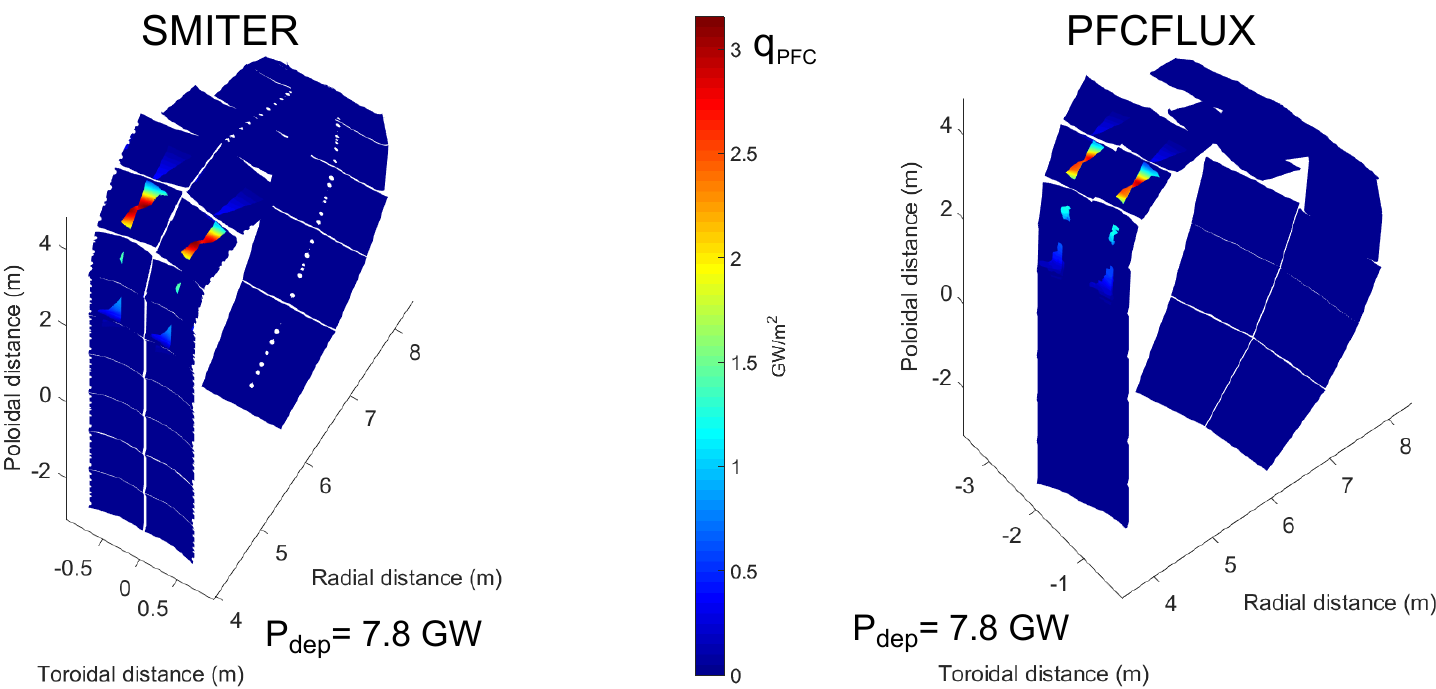}
  \caption{Comparison between the current quench thermal loads on a full 3-D poloidal ITER machine sector computed from SMITER and PFCFLUX~\cite{Lehnen}. The heat load distribution is
    derived assuming $P_{SOL}=8\,GW$ and $\lambda_q=30\,mm$. Due to different panel selection in toroidal direction the heat flux pattern on the left side in SMITER target setup is one half of the panel (wing) and corresponds to the right panel wing in PFCFLUX target setup and vice versa.
    }
  \label{fig:thermal_loads_comparison}
\end{figure*}
Fig.~\ref{fig:thermal_loads_comparison} compares the SMITER and PFCFLUX traces for the same magnetic equilibrium and input assumptions (see caption of Fig.~\ref{fig:thermal_loads_comparison}), although the geometry models used for the full poloidal sector differ in meshing. 
The target geometry in SMITER is at a different toroidal angle as compared to PFCFLUX. However, this has no impact on the heat transfer calculations due to the assumption of axisymmetric magnetic field.
PFCFLUX case setup uses complete panels, while SMITER uses two halves of the panels (wings) on the inboard side for target and complete panels on the outboard side, where mounting holes in the centre of the panels are clearly visible. 
Given these differences in the geometry, both the magnitude and wetted area distribution of the two tracers are similar and are in good agreement in terms of the magnitude of the deposited power density and the integrated value across all panels. 

The framework has also been utilized extensively for the development of real-time wall heat flux control algorithm for ITER~\cite{ANAND2018143}. Furthermore, the power flux density distribution estimated by the GUI framework has been successfully benchmarked against experimental IR diagnostic data on TCV tokamak~\cite{anand18:_validation_realtime_iter_tcv}.

\section*{Conclusion}

A new field-line tracing code SMITER has been developed primarily for ITER but can be used generally used on any tokamak. It allows power deposition mapping in the full 3-D CAD geometry of the machine taking as input a description of the radial profile of heat flow parallel to the magnetic field at the outer midplane of the  magnetic equilibrium.

It is embedded in a sophisticated, Python based GUI and incorporates state-of-the-art meshing capability, allowing the user to develop defeatured surface meshes of imported CAD models. Finite element thermal modelling tools are also being added allowing the computation of self-consistent temperature maps from the calculated surface power density distributions. The SMITER package is fully  integrated into the ITER Integrated Modelling Analysis Suite and is hosted (GIT version control) by the ITER Organization, available for use (and development) by all members of the ITER Parties. The code is already being actively employed for the development of simplified real time wall heat flux control algorithms~\cite{Anand} and is expected to play an important role in the production of synthetic diagnostic signals for the testing of ITER systems being prepared for PFC power flux control and monitoring.

\section*{Acknowledgements}

The work on SMITER was carried under contracts
ITER/RT/ 14/4300000958a1 and IO/16/CT/4300001401. The first
author has been partially supported for the MPI work that has been performed 
under the (HPC-EUROPA3 H2020 INFRAIA-2016-1-730897) 
programme. The views and opinions expressed herein do not necessarily
reflect those of the ITER Organization.


\end{document}